\documentclass{ws-ijmpcs}
\usepackage{slashed}
\usepackage{epstopdf}
\usepackage{color}
\begin{document}

\markboth{XIU-LEI REN et al.}
{OCTET BARYON MASSES ......}

%%%%%%%%%%%%%%%%%%%%% Publisher's Area please ignore %%%%%%%%%%%%%%%
%
\catchline{}{}{}{}{}
%
%%%%%%%%%%%%%%%%%%%%%%%%%%%%%%%%%%%%%%%%%%%%%%%%%%%%%%%%%%%%%%%%%%%%

\title{OCTET BARYON MASSES AND SIGMA TERMS IN COVARIANT BARYON CHIRAL PERTURBATION THEORY}

\author{XIU-LEI REN}

\address{School of Physics and
Nuclear Energy Engineering and
International Research Center for Nuclei and Particles in the Cosmos, Beihang University, Beijing, 100191, China\\
xiuleiren@phys.buaa.edu.cn}

\author{LI-SHENG GENG}

\address{School of Physics and
Nuclear Energy Engineering and
International Research Center for Nuclei and Particles in the Cosmos, Beihang University, Beijing, 100191, China\\
lisheng.geng@buaa.edu.cn}

\author{JIE MENG}

\address{School of Physics and
Nuclear Energy Engineering and
International Research Center for Nuclei and Particles in the Cosmos, Beihang University, Beijing, 100191, China\\
State Key Laboratory of Nuclear Physics and Technology, School of Physics, Peking University, Beijing, 100871, China\\
Department of Physics, University of Stellenbosch, Stellenbosch, 7602, South Africa\\
mengj@pku.edu.cn}

\maketitle

\begin{history}
\received{Day Month Year}
\revised{Day Month Year}
\end{history}

\begin{abstract}
We report an analysis of the octet baryon masses using the covariant baryon chiral perturbation theory up to next-to-next-to-next-to-leading order with and without the virtual decuplet contributions. Particular attention is paid to  the finite-volume corrections and the finite lattice spacing effects on the baryon masses. A reasonable  description of all the publicly available $n_f=2+1$ lattice QCD data is achieved.
Utilyzing the Feynman-Hellmann theorem, we determine the nucleon sigma terms as $\sigma_{\pi N}=55(1)(4)$ MeV and $\sigma_{sN}=27(27)(4)$ MeV.

\end{abstract}

\ccode{PACS numbers: 12.39.Fe,  12.38.Gc, 14.20.Gk}

\section{Introduction}	
Recently, the lowest-lying octet baryon masses have been studied on the lattice with $n_f=2+1$ configurations~\cite{Durr:2008zz,Alexandrou:2009qu,Aoki:2008sm,Aoki:2009ix,WalkerLoud:2008bp,Lin:2008pr,Bietenholz:2010jr,Bietenholz:2011qq,Beane:2011pc}. Because the limitation of the computational resources, most lattice quantum chromodynamics (LQCD) 
simulations still have to employ larger than physical light-quark masses, finite lattice volume and finite lattice spacing. Chiral perturbation theory (ChPT)~\cite{Weinberg:1978kz}, as an effective field theory of low-energy QCD, plays an important role in performing the multiple extrapolations needed
to extrapolate LQCD results (chiral extrapolations~\cite{Leinweber:2003dg,Bernard:2003rp,Procura:2003ig,Bernard:2005fy}, finite-volume corrections (FVCs)~\cite{Gasser:1986vb,Gasser:1987zq}, and continuum extrapolations~\cite{Beane:2003xv,Arndt:2004we}) to the physical world.

In this work we report on the first systematic study of the ground-state octet baryon masses in the covariant baryon chiral perturbation theory (BChPT) with the extended-on-mass-shell (EOMS) scheme up to next-to-next-to-next-to-leading order (N$^3$LO). The virtual decuplet contributions to the octet baryon masses and finite lattice volume and lattice spacing effects on the lattice data are studied. Finally, the octet baryon sigma terms are predicted using the Feynman-Hellmann theorem.

\section{Theoretical Framework}

Up to N$^3$LO, the octet baryon masses with the virtual decuplet contributions can be written as
\begin{equation}\label{Eq:N3LO}
  m_B = m_0 + m_B^{(2)} + m_B^{(3)} + m_B^{(4)} + m_B^{(D)},
\end{equation}
where $m_0$ is the chiral limit octet baryon mass, $m_B^{(2)}$, $m_B^{(3)}$, and $m_B^{(4)}$ correspond to the $\mathcal{O}(p^2)$, $\mathcal{O}(p^3)$, and $\mathcal{O}(p^4)$ contributions from the octet-only EOMS BChPT, respectively. The last term $m_B^{(D)}$ denotes the contributions of the virtual decuplet resonances up to N$^3$LO. Their explicit expressions and the corresponding FVCs can be found in Refs.~\cite{Ren:2012aj,Ren:2013dzt}.

In order to perform the continuum extrapolation of the LQCD simulations, one can first write down the Symanzik's effective filed theory~\cite{Symanzik:1983dc,Sheikholeslami:1985ij}. In Ref.~\cite{Ren:2013wxa}, we constructed the corresponding chiral Lagrangians up to $\mathcal{O}(a^2)$ to study the finite lattice spacing effects on the octet baryon masses, which can be written as
\begin{equation}\label{Eq:Aeffects}
  m_B^{(a)} = m_B^{\mathcal{O}(a)} + m_B^{\mathcal{O}(am_q)} + m_B^{\mathcal{O}(a^2)}.
\end{equation}

Here we want to mention that there are $19$ unknown LECs ($m_0$, $b_0$, $b_D$, $b_F$, $b_{1, \cdots,8}$, $d_{1,\cdots,5,7,8}$) needed to be fixed in the EOMS BChPT at $\mathcal{O}(p^4)$. Furthermore, including the finite lattice spacing effects (Eq.~(\ref{Eq:Aeffects})), one has to introduce $4$ more  combinations of the unknown LECs~\cite{Ren:2013wxa}.

\section{Results and Discussions}

The details of the studies can be found in Refs.~\cite{Ren:2012aj,Ren:2013dzt,Ren:2013wxa,Ren:2014vea}.
Here we only briefly summarize the main results.

In order to determine all the LECs and test the consistency of the current LQCD simulations, we perform a simultaneous fit to all the publicly available $n_f=2+1$ LQCD data from the PACS-CS~\cite{Aoki:2008sm}, LHPC~\cite{WalkerLoud:2008bp}, QCDSF-UKQCD~\cite{Bietenholz:2011qq},  HSC~\cite{Lin:2008pr}, and NPLQCD~\cite{Beane:2011pc} Collaborations. To ensure that the N$^3$LO BChPT stays in its applicability range, fitted LQCD data are limited to those satisfying $M^2_{\pi}< 0.25$ GeV$^2$ and $M_{\phi}L>4$.

In Refs.~\cite{Ren:2012aj,Ren:2013dzt}, we found that the octet-only EOMS BChPT shows a good description of the LQCD and experimental data with order-by-order improvement. Up to N$^3$LO, the $\chi^2/{\rm d.o.f.}$ is about $1.0$, which indicates that the lattice simulations from these five collaborations are consistent with each other~\footnote{This does not seem to be the case for the LQCD simulations of the ground-state decuplet baryon masses~\cite{Ren:2013oaa}.}, although their setups are very different. In addition, we showed that the explicit inclusion of the virtual decuplet baryons does not change the description of the LQCD data in any significant way, at least at $\mathcal{O}(p^4)$. This implies that using only the octet baryon mass data, one can not disentangle the virtual decuplet contributions from those of the virtual octet baryons and tree-level diagrams. On the other hand, we notice that the explicit inclusion of the virtual decuplet baryons does seem to improve slightly the description of the FVCs, especially for the LQCD data with small $M_{\phi}L$. Therefore, the virtual decuplet contributions to the octet baryon masses are not taken into account in our following studies.

To study discretization effects on the ground-state octet baryon masses, we constructed the
relevant chiral Lagrangians up to $\mathcal{O}(a^2)$ in Ref.~\cite{Ren:2013wxa}. 
By analyzing the latest $n_f=2+1$ $\mathcal{O}(a)$-improved LQCD data of the PACS-CS, QCDSF-UKQCD, HSC and NPLQCD Collaborations, we found that the finite lattice spacing effects are at the order of $1-2$\% for lattice spacings up to $0.15$ fm and the pion mass up to $500$ MeV, which is in agreement with other LQCD studies. 

Finally, the octet baryon sigma terms are predicted using the Feynman-Hellmann theorem. In order to obtain an accurate determination of sigma terms, a careful examination of the LQCD data is essential, since not all of them are of the same quality though they are largely consistent with each other. In Ref.~\cite{Ren:2014vea}, we only employed the PACS-CS, LHPC and QCDSF-UKQCD  data. We also took into account the scale setting effects of the LQCD simulations and studied systematic uncertainties from truncating chiral expansions. Furthermore, strong-interaction isospin breaking effects to the baryon masses were for the first time employed to better constrain the relevant LECs up to N$^3$LO. We predict the nucleon sigma terms as $\sigma_{\pi N}=55(1)(4)$ MeV and $\sigma_{sN}=27(27)(4)$ MeV, which are consistent with recent LQCD  and BChPT studies.

\section{Conclusions}

We have studied the lowest-lying octet baryon masses in the EOMS BChPT up to  N$^3$LO. The unknown low-energy constants are determined by a simultaneous fit to the latest $n_f=2+1$ LQCD simulations, and it is shown that the LQCD results are consistent with each other, though their setups are quite different. The  contributions of virtual decuplet resonances are explicitly included and we find that their effects on the octet baryon masses are small, especially for the chiral extrapolations.

We have studied finite-volume corrections and finite lattice spacing effects on the LQCD baryon masses as well. We find that their effects 
are of similar size  but finite volume corrections are more important to better constrain the LECs and to reduce the $\chi^2/\mathrm{d.o.f.}$.

Using the Feynman-Hellmann theorem, we have performed an accurate determination of the nucleon sigma terms, focusing on the uncertainties from the lattice scale setting method and chiral expansions. Our predictions are $\sigma_{\pi N}=55(1)(4)$ MeV and $\sigma_{sN}=27(27)(4)$ MeV, which are
 consistent with most of the recent LQCD and BChPT studies. However, further LQCD simulations
 are needed to reduce the uncertainty of the nucleon strangeness-sigma term.

\section*{Acknowledgments}
X.-L.R acknowledges the Innovation Foundation of Beihang University for Ph.D. Graduates.
This work was partly supported by the National
Natural Science Foundation of China under Grants No. 11005007,  No. 11375024, and No. 11175002, and the New Century Excellent Talents in University  Program of Ministry of Education of China under Grant No. NCET-10-0029,  the Fundamental Research Funds for the Central Universities, and
the Research Fund  for the Doctoral Program of Higher Education under Grant No. 20110001110087.

%\begin{thebibliography}{000} %for 3 digits
%\begin{thebibliography}{00}  %for 2 digits

\end{document}